\documentclass[twocolumn,showpacs,preprintnumbers,amsmath,amssymb,aps,prb,superscriptaddress]{revtex4}
\usepackage{graphicx}
\usepackage[latin1]{inputenc}
\newcommand{\ket}[1]{\left\vert{#1}\right\rangle}
\newcommand{\bra}[1]{\left\langle{#1}\right\vert}

\newcommand{\moy}[1]{\left\langle{#1}\right\rangle}

\def\bea{\begin{eqnarray*}}
\def\eea{\end{eqnarray*}}
\def\be{\begin{equation*}}
\def\ee{\end{equation*}}
\def\Ham{\mathcal{H}}
\def\Struc{\mathcal{S}}
\def\P{\mathcal{P}}
\def\k{\mathbf{k}}
\def\r{\mathbf{r}}
\def\h{\textrm{h}}
\def\S{\mathbf{S}}

\def\fig{FIG. }
\begin{document}

\author{G.\ Roux} \affiliation{Laboratoire de Physique Th\'eorique,
CNRS UPS UMR-5152, F-31062 Toulouse, France.}  \author{S.\ R.\ White}
\affiliation{Department of Physics, University of California, Irvine
CA 92697, USA.} \author{S.\ Capponi} \affiliation{Laboratoire de
Physique Th\'eorique, CNRS UPS UMR-5152, F-31062 Toulouse, France.}
\author{A.\ L\"auchli} \affiliation{Institut Romand de Recherche
Num\'erique en Physique des Mat\'eriaux, CH-1015 Lausanne,
Switzerland.} \author{D.\ Poilblanc} \affiliation{Laboratoire de
Physique Th\'eorique, CNRS UPS UMR-5152, F-31062 Toulouse, France.}
\affiliation{Institute of Theoretical Physics, EPFL, BSP, CH-1015
Lausanne, Switzerland.}

\date{\today} 

\title{Doped two-leg ladder with ring exchange : exact diagonalization
    and density matrix renormalization group computation}


\begin{abstract}
The effect of a ring exchange on doped two-leg ladders is investigated
combining exact diagonalization (ED) and density matrix
renormalization group (DMRG) computations. We focus on the nature and
weights of the low energy magnetic excitations and on superconducting
pairing. The stability with respect to this cyclic term of a
remarkable resonant mode originating from a hole pair - magnon bound
state is examined. We also find that, near the zero-doping critical
point separating rung-singlet and dimerized phases, doping reopens a
spin gap.
\end{abstract}
\maketitle

Two-leg cuprate ladders~\cite{Dagotto1996} and two-dimensional (2D)
high-temperature (high-T$_c$) superconductors seem to surprisingly
share many similarities. In addition, ladders exhibit exotic physical
behaviors characteristic of one-dimensional systems. Under high
hydrostatic pressure, the intrinsically doped
Sr$_2$Ca$_{12}$Cu$_{24}$O$_{41+\delta}$ ladder
material~\cite{ladder:supercond2} becomes superconducting hence
reinforcing the similarity between two-leg ladder and layer-based
cuprates. Doped two-leg ladders are also appealing for
theorists~\cite{Dagotto1996}~: its ground state (GS) is believed to be
a simple realization of the Resonating Valence Bond (RVB) state
proposed by Anderson~\cite{RVB} in the context of high-T$_c$. In
addition, ladders are one of the simplest realistic correlated
electron systems to investigate novel mechanisms of
superconductivity~\cite{PRL2004}. Interestingly, new low-energy
magnetic excitations have been observed by Nuclear Magnetic Resonance
(NMR) under pressure, i.e under doping the ladder
planes~\cite{ladder:supercond2}, a finding possibly connected to
predicted exotic magnon-hole pair states~\cite{Poilblanc2000}.

Taking into account ring exchange has proven to be essential to
understand the spin dynamics of insulating cuprate materials such as
both ladders~\cite{Matsuda2000} and high-T$_c$~\cite{Coldea2001}. Ring
exchange which permutes cyclically spins on square plaquettes is known
to suppress magnetic order in several 2D systems. It naturally appears
in a fourth order expansion in $t/U$ of the Hubbard model near
half-filling and also in \emph{ab initio}
calculations~\cite{Calzado2003}. In order to study the properties of
lightly doped ladders~\cite{Hayward1995} with ring exchange, we shall
use the framework of the $t-J$ Hamiltonian and include a cyclic
exchange of magnitude $K$ \bea \Ham &=& -t \sum_{<ij>,\sigma}
\P_G[c_{i\sigma}^{\dag} c_{j\sigma}
+c_{j\sigma}^{\dag}c_{i\sigma}]\P_G\\ &&+ J
\sum_{<ij>}[\S_{i}\cdot\S_{j}-\frac{1}{4}n_i n_j] +\,K\sum_{ijkl\in
\square}[P^{\circlearrowleft}_{ijkl} + P^{\circlearrowright}_{ijkl}]
\eea with $\P_G$ Gutzwiller projectors. Labels $i,j,k,l$ of
permutation operators $P^{\circlearrowleft, \circlearrowright}_{ijkl}$
stand for the sites encountered along each square plaquette. The
expression of the cyclic exchange in terms of spin $1/2$ operators
\bea P^{\circlearrowleft}_{ijkl} + P^{\circlearrowright}_{ijkl} & = &
\frac{1}{4}\!+\! \left[\S_{i}\cdot\S_{j} + \S_{j}\cdot\S_{k} +
\S_{k}\cdot\S_{l} + \S_{l}\cdot\S_{i}\right] \\ & + & \left[
\S_{i}\cdot\S_{k} + \S_{j}\cdot\S_{l} \right] + \,4\left[
(\S_{i}\cdot\S_{j})(\S_{k}\cdot\S_{l})\right. \\ && \left. +
(\S_{i}\cdot\S_{l})(\S_{j}\cdot\S_{k}) -
(\S_{i}\cdot\S_{k})(\S_{j}\cdot\S_{l}) \right] \eea includes bilinear
frustrating terms and biquadratic terms~\cite{Muller2002}. The $K$
term leads to a rich phase diagram at
half-filling~\cite{Lauchli2003}. In particular, a phase transition
between the RVB rung-singlet phase and a phase with staggered dimer
order occurs~\cite{Muller2002, Lauchli2003, Nersesyan1997, Hijii2002}
at the critical point $K_c/J \simeq 0.2$. Experimentally
proposed~\cite{Matsuda2000} values of $K/J$ for the insulator ladder
material range in a window $0.025\leq K/J\leq 0.075$. We are also
interested here in the proximity of this critical point. Thus, we
shall restrict ourselves to parameters $-0.2\leq K/J\leq 0.2$ and
assume a physical value of $J/t=0.5$ for which superconducting
fluctuations are dominant at low doping~\cite{Hayward1995} when
$K/J=0$. A similar model has recently been studied on the 2D square
lattice~\cite{Chung2003} in a large $N$ limit of a derived $SU(N)$
model.

It is instructive to consider in more detail the effect of the cyclic
exchange in terms of the RVB description of ladders~\cite{White1994}.
Consider a single plaquette. With $K=0$, the ground state is exactly
described as an RVB state, with the horizontal dimer state added to
the vertical dimer state~\cite{Footnote1}. A small nonzero $K$ rotates
between horizontal and vertical valence bond states and acts either to
enhance ($K<0$) or suppress ($K>0$) the resonance. The ground state
has energy $E= -3J + 2K$. For $K=J/2$, the resonance is destroyed and
the singlet ground state is doubly degenerate between linear
combinations of the two VB states. On the other hand, a nonzero $K$ of
either sign is frustrating to N\'eel order. Thus the cyclic term
mediates between VB and RVB order.

\begin{figure}[t]
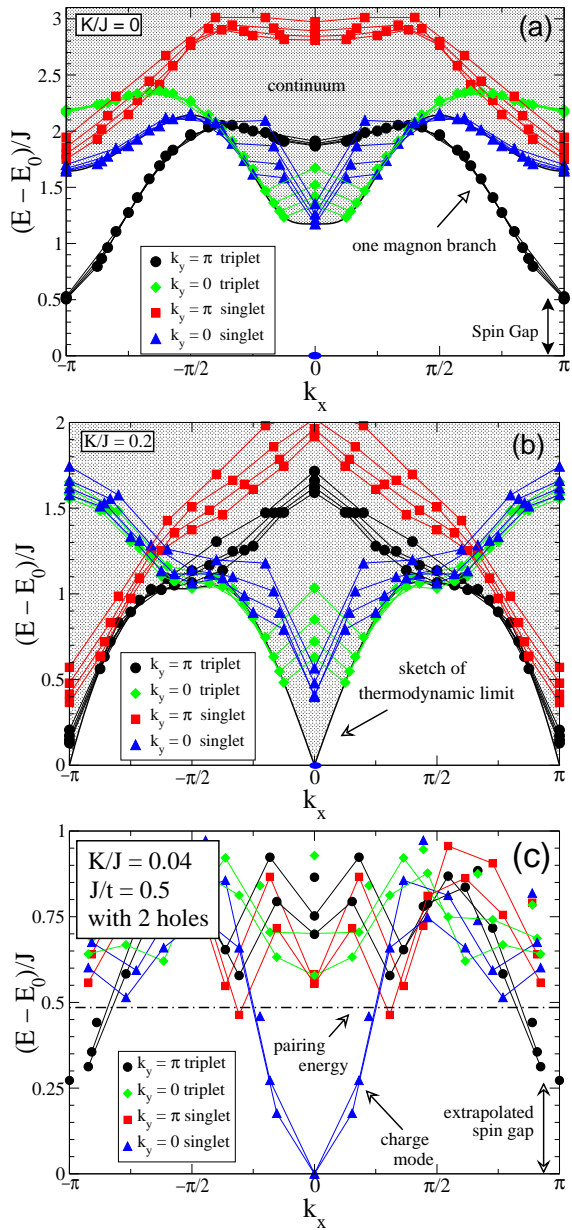

\includegraphics[width=5.4cm,angle=270,clip]{fig1.eps}
\includegraphics[width=5.4cm,angle=270,clip]{fig2.eps}
\includegraphics[width=5.4cm,angle=270,clip]{fig3.eps}
\caption{(Color online) \textbf{(a,b)} Spectra for undoped ladders
from ED calculations on ladders of length $L=12,14$ and 16. Grey areas
are guide to the eyes. \textbf{(c)} Spectra for ladders doped with two
holes and $L=9,11,13$. Note that $k_x=\pi$ does not belong to the
Brillouin zone of systems with odd $L$. The dashed line shows
$\Delta_p$ computed independently.}
\label{spectra}
\end{figure}
In this article, after briefly reviewing the undoped case in
connection with previous work, we first consider the case of two holes
in a ladder, a simple limit where we expect to get physical insight on
the various elementary excitations carrying different spin and charge
quantum numbers. In a second step, we address the finite doping issue
where we identify the previous excitations, discuss the effect of
doping and connections to experiments.

\emph{Elementary excitations} -- The magnetic energy spectrum is shown
in \fig \ref{spectra} (a). One can distinguish a one magnon branch
excitation around its minimum at momentum $ \k = (\pi,\pi)$. The
bottom of a two-magnon continuum is located around $\k = (0,0)$ with
an energy about twice the spin gap. When $K/J$ increases, the spin gap
is reduced and vanishes at the critical point as was previously
proposed. There (\fig \ref{spectra} (b)), the triplet branch is
divided into a symmetric mode ($k_y = 0$, $k_x \in [0,\pi/2]$)
reminiscent of the two-magnon continuum, and an antisymmetric mode
($k_y = \pi$, $k_x \in [\pi/2,\pi]$) reminiscent of the one-magnon
branch. The doped system (\fig \ref{spectra} (c)) is a Luther-Emery
liquid provided $K/J$ remains small enough : a gapless charge mode
appears at low momenta, corresponding to the motion of the hole pair,
while the one magnon branch remains gapped. A quasiparticle (QP)
continuum starts for an energy which equals the pairing energy
(computed independently, see after) and larger than the spin gap. Note
that the triplet branch has its minimum at an incommensurate momentum
$\k_{\delta} = (\pi (1-\delta),\pi)$ and that, in the case of two
holes under study, $\k_{\delta} \rightarrow (\pi,\pi)$ in the
thermodynamic limit. As $K/J$ increases, the spin gap is reduced as is
the onset of the QP continuum. Lastly, the system with $K<0$ has
similar behavior to a two-leg ladder with a rung coupling $J_{\perp} >
J_{\parallel}$.

In order to distinguish between the collective magnon and the QP
excitations, we first focus on ladders with 2 holes forming a singlet
$d$-wave bound state when $L \rightarrow \infty$. Following ref. 6, we
define \bea \Delta_M&=&E(0\h,S=1)-E(0\h,S=0)\\
\Delta_S&=&E(2\h,S=1)-E(2\h,S=0)\\
\Delta_p&=&2E(1\h,S=1/2)-E(2\h,S=0)-E(0\h,S=0)\,, \label{energies}
\eea where $E(n\h,S)$ is the GS energy with $n$ holes and spin
$S$. $\Delta_M$ represents the magnon gap, $\Delta_S$ the spin gap
with two holes, and $\Delta_p$ the pairing energy.
\begin{figure}
\includegraphics[width=5.7cm,angle=270,clip]{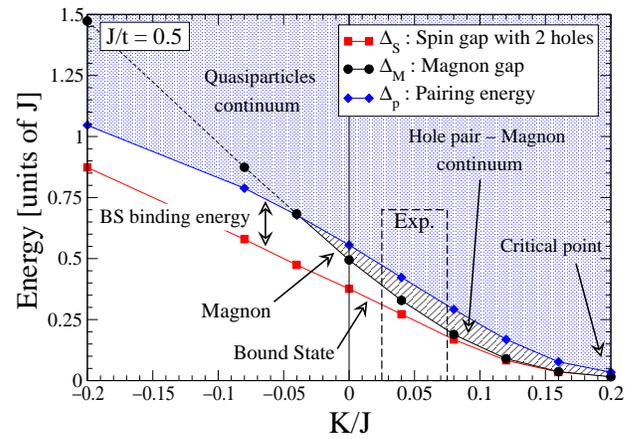}
\caption{(Color online) Comparison between characteristic energies of
elementary magnetic excitations in doped ladders. Data are
extrapolated from DMRG computations on systems of length up to
$L=48$. Proposed experimental values of $K/J$ for undoped systems are
in the Exp. window.}
\label{comparison}
\end{figure}
Each gap shown on \fig \ref{comparison} has been calculated on finite
systems with DMRG~\cite{dmrg} and then extrapolated to the thermodynamic limit.
First, let us make a few general statements on these energies : (i)
since a magnon can be created arbitrarily ``far away'' from the holes,
then $\Delta_S \le \Delta_M$, (ii) breaking the hole pair creates two
quasiparticles so that $\Delta_S \le \Delta_p$. These inequalities are
well verified in the data. The discontinuity of the spin gap in the
infinitesimal doping limit has been attributed to the existence of a
bound state between the hole pair and the magnon, responsible for a
sharp resonating mode at finite density similar to what can be
observed in Inelastic Neutron Scattering (INS) experiments on
underdoped cuprates. The binding energy of this bound state can be
defined as $\min(\Delta_p,\Delta_M) - \Delta_S$, see
\fig\ref{comparison}. When $\Delta_p < \Delta_M$, the continuum starts
at the QP continuum so that the magnon is scattered by QP and acquires
a finite life time (dashed part of the $\Delta_M$ line, $K/J \leq
-0.04$). When $\Delta_M < \Delta_p$, we expect a continuum of hole
pair-magnon scattering states above $\Delta_M$ and below the QP
continuum, as shown by the grey dashed area for $-0.04\leq K/J\leq
0.2$. The binding energy of the bound state is strongly affected by
the cyclic exchange with an instability treshold around a value of $K$
significantly smaller than the one of the critical point. Moreover,
the pairing energy vanishes simultaneously with the spin gap when $K$
reaches $K_c$.

\begin{figure}
\includegraphics[width=3.2cm,angle=270]{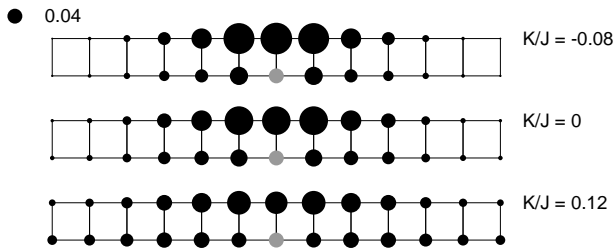}
\caption{Hole-hole static correlation functions from ED computations
on $L=$ 13 ladder with two holes for different $K/J$. The area of
circles is proportional to the correlation value. Up left circle
indicates the mean expectation value.}
\label{corr}
\end{figure}

\begin{figure}
\includegraphics[width=3.5cm,angle=270]{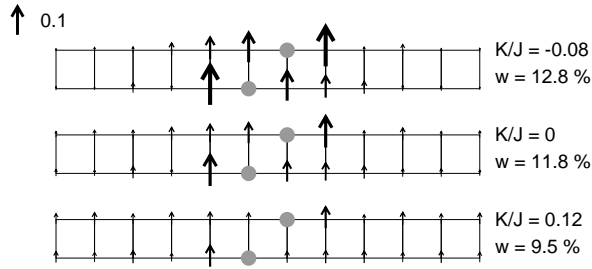}
\caption{Spin density around the hole pair in the first triplet state
from ED data on $L=$ 13 ladder with two holes for different
$K/J$. Weights \texttt{w} of the projected function are
indicated. Lengths of arrows are proportional to the local density
values.}
\label{corrBS}
\end{figure}

The binding of holes (and also spinons) in the two-leg ladder can be
described as an effect of the resonance in the RVB
language~\cite{White1994} : an isolated hole induces a staggered,
non-resonating VB state either to the left or the right of the hole,
which is only healed by another hole. The destruction of resonance as
$K$ is increased towards $K/J\approx 0.2$ should translate directly
into weakened and eventual loss of pairing.  This effect is clearly
evident in \fig \ref{corr}. For $K<0$, the pairing is strengthened
both by the enhancement of resonance and by the fact that the cyclic
exchange does not act on a plaquette with a hole, favoring states with
holes sharing plaquettes.

The pair-magnon binding can be viewed as closely related to
ferromagnetic spin polarons. The hopping of holes in a ferromagnetic
background is unfrustrated; the magnon produces a small region of
enhanced ferromagnetic correlations. In \fig\ref{corrBS}, the bound
state has been projected onto the states with two holes on the
diagonal of a plaquette (of weights \texttt{w} given on figure)
showing indeed a strong ferromagnetic region for $K<0$.  More
precisely, the hole pair--magnon binding results from a gain in
kinetic energy coming from processes that exchange the triplet and the
pair locally on a plaquette~\cite{Jurecka2002}.  As $K/J$ increases,
the ``free'' magnon excitation lowers its energy while the above
resonant processes are less affected.

\begin{figure}
\includegraphics[width=5.4cm,angle=270,clip]{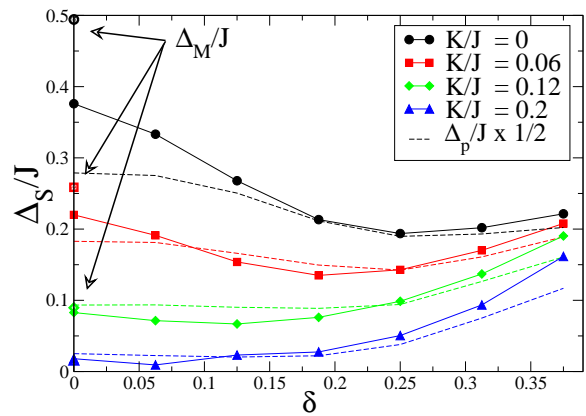}
\caption{(Color online) Spin gap for doped two-leg ladders as a
function of hole doping $\delta$. Data extrapolated from DMRG
computations on systems of length $L =$ 16 to 64 and doping $\delta =
n/16$, $n = 0,\cdots,6$, with a number of kept states up to
$m=1600$. The $\delta=0$ magnon energies are indicated by grey symbols
(see arrows). \emph{Half} the pairing energy for a finite system is
also given by dashed lines.}
\label{density}
\end{figure}

\emph{Finite doping study} -- We now turn to the more complex case of
finite doping. As shown below, we find signatures of the various
excitations identified in the previous part. The spin gap for several
hole density $\delta$ is displayed in \fig \ref{density}. For $K/J=0$,
we see that it is reduced under doping because holes weaken spin
exchange~\cite{Jeckelmann1998} but remains
robust~\cite{Poilblanc2004}. For small magnitude of cyclic exchange
(for instance $K/J=0.06$) this effect is still significant, leading to
an appreciable lowering in comparison with the undoped system. As the
critical point is approached, the magnetic parent acquires a smaller
spin gap but the relative decrease under doping is less and less
significant. Interestingly, at the half-filling critical point $K_c/J
\simeq 0.2$, doping eventually leads to a reopening of the spin
gap. Qualitatively, this could be explained by the fact that a hole on
a plaquette prevents the action of the ring exchange. Therefore, at
sufficiently large doping ($\delta\simeq 0.4$), the spin gap depends
weakly on $K$ for studied values. Pairing energy also increases at
large densities which suggests a deep relation between pairing and
spin gap.

\begin{figure}
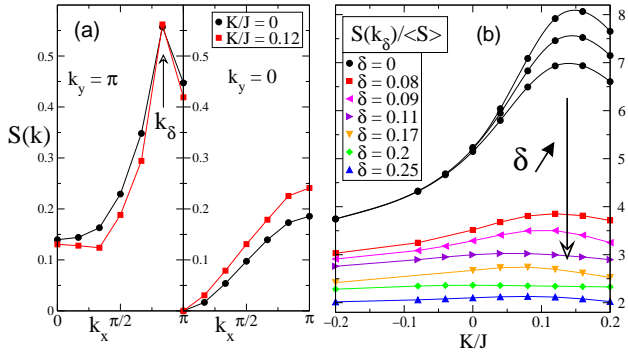

\includegraphics[width=4.5cm,angle=270,clip]{fig8.eps}
\includegraphics[width=4.6cm,angle=270,clip]{fig9.eps}
\caption{(Color online) \textbf{(a)} static structure factor
$\Struc(\k)$ for a ladder of length 12 with 4 holes ($\delta =
0.17$). \textbf{(b)} $\Struc(\k_{\delta}) / \moy{\Struc}$ for
different doping as a function of $K/J$, from systems with $L$ ranging
from 8 to 16. $\Struc(\k_{\delta})$ is normalized w.r.t. the average
structure factor $\moy{\Struc} = (2L)^{-1}\sum_{\k}\Struc(\k) =
\frac{1}{4} (1-\delta)$.} \label{statics}
\end{figure}

\begin{figure}[t]
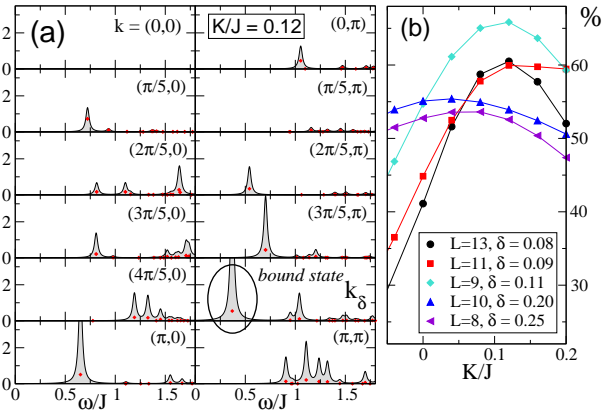

\includegraphics[width=5.4cm,angle=270,clip]{fig10.eps}
\includegraphics[width=5cm,angle=270,clip]{fig11.eps}
\caption{(Color online) \textbf{(a)} dynamic structure factor
$\Struc(\k,\omega)$, with $\k=(k_x,k_y)$, for a ladder of length 10
with 4 holes. Red dots indicate poles and their relative
weight (see text). \textbf{(b)} relative weight of the resonant state
at $\k_{\delta}$.} \label{dynamics}
\end{figure}

We have computed static and dynamic spin structure factors for undoped
and doped systems by ED with standard continued-fraction techniques
and studied their evolution with increasing $K/J$. The static
structure factor $\Struc(\k)$ gives insights on possible magnetic
ordering. The dynamic spin structure factor $\Struc(\k,\omega)$ probes
magnetic excitations that can be detected in INS experiments.  We use
the standard definitions \bea \Struc(\k)&=& \moy{S^z_{-\k}S^z_{\k}}_0
= \int d\r e^{i\k \cdot \r}\moy{S^z(0)S^z(\r)}_0\,. \\
\Struc(\k,\omega) &=& -\frac{1}{\pi}\Im \bra{\psi_0} S^z_{-\k}
\frac{1}{\omega+E_0+i\eta-\Ham} S^z_{\k} \ket{\psi_0} \,, \eea where
$E_0$ is the ground state energy and $\eta$ gives a finite broadening
to the peaks. Within our choice of normalization,
$\Struc(\k_{\delta})$ should either diverge as $L^{\alpha}, \alpha >
0$ or $\ln L$ for quasi-ordered states or converge for states with
fast decaying correlations. At fixed $L$, $\Struc(\k_{\delta})$
follows the finite correlation length and shows a maximum defining a
critical point $K_c(L)$. In the thermodynamic limit,
$K_c(L)\rightarrow K_c$. Such a behaviour is clear at half-filling
(see \fig \ref{statics} (b), black circles corresponding to $L=12,14$
and $16$). From the data at finite density, the incommensurate nature
of the resonant mode is clearly seen and consistent with the expected
value $\k_{\delta}$. This feature is preserved up to the critical
point. Moreover, in contrast to the undoped case, the maximum of
$\Struc(\k_{\delta})$ with $K$ is hardly noticeable at large
density. This is consistent with the finite value of the spin gap
found at large doping.

Inspecting $\Struc(\k,\omega)$ reveals a pole well separated from the
rest of the spectrum at momentum $\k=\k_\delta$ (\fig \ref{dynamics}
(a)).  This peak can be clearly identified as the resonant mode issued
from the triplet BS discussed above~\cite{Poilblanc2000}. Although ED
suffers from finite size effects and dynamic spectra are tedious to
analyze due to multi-excitation continua, the evolution of its
relative weight given by $|\bra{\psi_{\textrm{Res}}} S^z_{\k}
\ket{\psi_0}|^2 / \Struc(\k_{\delta})$ (where
$\ket{\psi_{\textrm{Res}}}$ is the resonant state wave function) is
still very instructive. Note that in the thermodynamic limit, the
resonant mode should not give rise to a single pole but rather a
singularity \cite{Poilblanc2004}. The evolution of the relative weight
is displayed in \fig \ref{dynamics} (b) and shows that the BS  remains
robust at finite density and $K>0$, keeping a sizeable fraction of the
total weight. Consequently, we predict that it should be observable in
spin dynamics sensitive experiments. This conclusion on the robustness
and observability of the Luther-Emery phase is in agreement with
recent NMR experiments on doped ladder materials\cite{Fujiwara2003}.

\emph{Conclusion} -- In summary, we show that the nature of the
Luther-Emery phase and its remarkable resonant mode is preserved for
intermediate physical $K/J$ values. In this regime, both spin gap and
superconducting pairing energy are very sensitive to the ring
exchange. In addition, doping is shown to have a drastic effect at the
magnetic critical point of the parent Mott insulator.

\emph{Acknowledgments} -- G.\ R. would like to thank IDRIS (Orsay,
France) for use of supercomputer facilities. S.\ R.\ W. acknowledges
the support of the NSF under grant DMR03-11843. D.\ P. thanks the
Institute for Theoretical Physics (EPFL, Switzerland) for hospitality.



\begin{thebibliography}{23}
\bibitem{Footnote1} Note that the sign between these two VB terms is
negative when the state is expressed in fermionic representation; this
minus sign is closely tied to the d-wave nature of pairing in doped
antiferromagnets.
\end{thebibliography}
\end{document}